\documentclass[11pt]{article}

\usepackage{amsmath}   
\usepackage{amssymb}
\usepackage{amsfonts}

\usepackage[margin=2cm]{geometry}

\DeclareMathOperator  \Tr {Tr}

\title{Flavour Symmetric Mass Matrices}
\author{A. Kleppe\\
SACD, Oslo, Norway}
\date{}

\begin{document}

\maketitle
\begin{abstract}
The structure of flavour space is determined by the form of the quark mass matrices in the weak flavour space basis. We examine some matrix textures in the light of flavour permutations symmetry arguments, for three and four families.
\end{abstract}


\section{Mass matrices in flavour space}

The fermion masses live in flavour space. Mixing, as well as CP-violation, are flavour space phenomena, and
the structure of flavour space is determined by the mass 
matrices, i.e. by the form that the mass matrices take in the weak basis in flavour space. This is the basis where the mixed fermion states interact weakly, in contrast to the mass bases, where the mass matrices are diagonal, $diag(m_1,m_2,m_3)$, $m_j$ being the masses of the physical fermions.
 
The information content of a $N\times N$ matrix $M$ is contained in its $N$ matrix invariants, which are the sums, as well as the sums of products, of the eigenvalues, like $traceM$, $detM$, and so on,
\[ 
\lambda_1+\lambda_2+\lambda_3... = \sum_j\lambda_j, 
\]
\[ 
\lambda_1\lambda_2+\lambda_1\lambda_3+\lambda_1\lambda_4+... = \sum_{jk}\lambda_j\lambda_k,
\]
\[ 
\lambda_1\lambda_2\lambda_3+\lambda_1\lambda_2\lambda_4+...= \sum_{jkl}\lambda_j\lambda_k\lambda_l,
\]
where the indices ${j,k,l,...}$ run over all the $N$ eigenvalues.
These expressions are invariant under permutations of the eigenvalues, and thereby of the eigenstates, which in the context of mass matrices means that they are flavour symmetric, i.e. invariant under permutations of the flavours, as well as under unitary transformations of the mass matrices.

The invariants of a given finite matrix can always be calculated even if the eigenvalues are not known, i.e. the $N$ invariants of a $N \times N$ matrix $M$are  
\[
\Tr(M)= m_1+m_2+m_3+...+m_N
\]
\[
m_1m_2+m_1m_3+m_2m_3+...m_{N-1}m_N = \frac{1}{2}[(\Tr M)^2 - \Tr(M^2)]
\]
\[
m_1m_2m_3+m_1m_2m_4+...m_{N-2}m_{N-1}m_N = \frac{1}{6}[(\Tr
M)^3+2\Tr(M^3)-3 \Tr M \Tr(M^2)]
\] 
$
m_1m_2m_3m_4+m_1m_2m_3m_5+...+m_{N-3}m_{N-2}m_{N-1}m_N = \frac{1}{24}[(\Tr{\bf{M}})^4+3(\Tr({\bf{M}}^2))^2$
$+8(\Tr({\bf{M}}^3))\Tr({\bf{M}})-6\Tr({\bf{M}}^4)-6\Tr({\bf{M}}^2)(\Tr{\bf{M}})^2]
$, 
{\vspace{3mm}}

\noindent and so on. Even though the matrix information is contained in its invariants, it is the form that the mass matrices take in the weak basis that carries the information about the flavour space structure, and can give some hint about the family hierarchical patterns. On the one hand there is the mass hierarchy, where the masses of the third families completely dominate, and the second family masses in their turn are much larger than the masses of the first families. The question is why fermions of the same charge but in different families, in spite of having identical couplings to all gauge bosons of the strong, weak and 
electromagnetic interactions, nevertheless exhibit such divergent mass values,
ranging from the electron mass to the about $10^{5}$ times larger top mass.
On the other hand, there is the hierarchy exhibited by the mixing matrix, where the transitions between the first and second family are far greater than the transitions between the second and third family and even more so, between the first and the third family. 

There is no obvious principle or pattern relating these masses, and in the hope to find a mass texture that could shed some light on the origin of the mass spectra, an industry of different mass matrix ans{\"a}tze has emerged - 
to make an ansatz for a quark mass matrix basically amounts to guessing what structure the matrix has in the weak flavour space basis.

\section{Permutation symmetries}
 
In the Standard Model, where fermions get their masses from the Yukawa 
couplings via the Higgs mechanism, there is no reason why there should 
be a different Yukawa coupling for each fermion, and it seems  
natural that these couplings would all be similar. 
In the ``democratic scheme''\cite{ak1demo}, \cite{ak1koide}, \cite{ak1Fritzsch} this absence of any distinguishing principle is taken at face value, and with the assumption that the Yukawa couplings of the fermions 
of a given charge are equal, a ``zeroth order mass matrix'' is postulated, with the form $M_0$ = $k{\bf{N}}$, where $k$ has dimension mass, and ${\bf{N}}$ is the $S(3)_L  \times  S(3)_R $ flavour symmetric Nambu matrix 
\begin{equation}\label{ak1nambu}
{\bf{N}} = \left (\begin{array}{rcl}
                  1   &  1{\hspace{4mm}}1\nonumber\\
                  1   &  1{\hspace{4mm}}1\nonumber\\
                  1   &  1{\hspace{4mm}}1
                      \end{array}
                \right).
\end{equation}
The Nambu matrix has the eigenvalue spectrum $(0,0,3)$, reflecting
the charged fermion mass spectra with two light families and a third much heavier family, a mass hierarchy that can be interpreted as the repesentation ${\bf{1}}\oplus {\bf{2}}$ of $S(3)$. 
In order to obtain realistic mass spectra with non-zero masses, the $S(3)_L  \times  S(3)_R $ flavour symmetry must be broken, 
and different schemes for symmetry breaking correspond to different matrix ans\"{a}tze. 

Any scheme for breaking the democratic symmetry amounts to perturbing
the democratic matrix, i.e. 
\begin{equation}\label{ak1dimi}
   M_0 \rightarrow M= 
            k  \left(\begin{array}{rcl}
                  1   &  1{\hspace{4mm}}1\nonumber\\
                  1   &  1{\hspace{4mm}}1\\
                  1   &  1{\hspace{4mm}}1\nonumber
                      \end{array}
               \right) + {\bf{\Lambda}},
\end{equation}
where $k$ has dimension mass, and ${\bf{\Lambda}}$ is a 3x3 matrix which is much "smaller" than $M_0$.

The actual observed mass spectra, with two very light masses, and a heavy third family fermion,  agree with this pattern, and be interpreted as a remnant of a (broken) $S(2)_L \times S(2)_R $ symmetry.

Taking this into account, it is natural to apply a stepwise breaking scheme on the initial $S(3)_L \times S(3)_R $ 
symmetry, first breaking down the $S(3)_L \times S(3)_R $ to
$S(2)_L \times S(2)_R $, and so on, in order to obtain the physical mass matrix.

The most general way of breaking $S(3)_L \times S(3)_R $ to $S(2)_L \times S(2)_R $ corresponds to the matrix  
\begin{equation}\label{ak1namba}
{\bf{M}} = \left(\begin{array}{rcl}
                  A   &  A{\hspace{4mm}}B\nonumber\\
                  A   &  A{\hspace{4mm}}B\nonumber\\
                  B   &  B{\hspace{4mm}}C
                      \end{array}
                \right),
\end{equation}
with the eigenvalues 
$$(0,[2A+C-\sqrt{(2A-C)²+8B^2}]/2,[2A+C+\sqrt{(2A-C)²+8B^2}]/2).$$
We however desire a mass spectrum with massive states, and a $S(2)_L \times S(2)_R $ flavour symmetric mass matrix with only non-zero mass values and akin to ({\ref{ak1namba}}), is
\begin{equation}\label{ak1nambi}
{\bf{K}} = \left(\begin{array}{rcl}
                  A   &  B{\hspace{4mm}}B\nonumber\\
                  B   &  A{\hspace{4mm}}B\nonumber\\
                  B   &  B{\hspace{4mm}}A
                      \end{array}
                \right),
\end{equation}
with the mass spectrum $(A-B,A-B,A+2B)$, and $A = (m_1+m_2+m_3)/3$, $B= (2m_3-m_2-m_1)/6$.
Just like the Nambu matrix, ${\bf{K}}$ has two degenerate mass values, but unlike the Nambu mass spectrum, where the 
third family is dramatically singled out, the eigenvalues of the mass matrix 
${\bf{K}}$ may all be of the same order of
magnitude, in the case $A \gg B$. 

This type of mass matrix does not reflect the spectra of the charged fermions, but could be relevant for massive neutrinos. Data on neutrino masses 
indicate that non-vanishing neutrino masses do not follow the same pattern as the
charged fermion mass spectra, since unlike the other fermion masses, the neutrino masses 
seem to be rather aligned. This gives a suspicion of a flavour symmetry which is
merely slightly broken also in the mass basis.

\section{The charged fermion mass matrices}
In the case of the charged fermions, the flavour symmetry must however be radically broken.
Starting with the matrix ${\bf{K}}$, 
there are several possible ways to stepwise break its
flavour symmetry. One variation of
${\bf{K}}$ is
\begin{equation}\label{ak1nambo}
{\bf{K'}} = \left(\begin{array}{rcl}
                  A   &  B{\hspace{4mm}}D\nonumber\\
                  B   &  A{\hspace{4mm}}D\nonumber\\
                  D   &  D{\hspace{4mm}}A
                      \end{array}
                \right)
\end{equation}
with the mass spectrum $(A-B,[2A+B-\sqrt{B^2+8D^2}]/2,[2A+B+\sqrt{B^2+8D^2}]/2)$ which is symmetric in the first two flavours in the weak basis. Now consider the overdetermined matrix
\begin{equation}\label{ak1namby}
{\bf{L}} = \left(\begin{array}{rcl}
                  A   &  B{\hspace{4mm}}D\nonumber\\
                  B   &  A{\hspace{4mm}}D\nonumber\\
                  D   &  D{\hspace{4mm}}C
                      \end{array}
                \right)
\end{equation}
with the mass spectrum 
$(A-B,[A+B+C-\sqrt{(A+B-C)^2+8D^2}]/2$,$[A+B+C+\sqrt{(A+B-C)^2+8D^2}]/2)$.

If ${\bf{L}}$ is the up-sector mass matrix, we assume that the down-sector mass matrix of the same texture,
\begin{equation}\label{ak1nambyd}
{\bf{L'}} = \left(\begin{array}{rcl}
                  A'   &  B'{\hspace{4mm}}D'\nonumber\\
                  B'   &  A'{\hspace{4mm}}D'\nonumber\\
                  D'   &  D'{\hspace{4mm}}C'
                      \end{array}
                \right)
\end{equation}

Different values for the redundant parameter $D$ leads to different mass textures.    
The choice $D = 0$ corresponds to the eigenvalues 
$(A-B,[A+B+C\pm (A+B-C)]/2)$, i. e. $(m_1,m_2,m_3) = (A-B, A+B, C)$,
and correspondingly for the down sector, and $D = C - A - B$ to the eigenvalues $(A - B, 2(A+B)-C, 2C - (A+B))$.

\noindent We can numerically evaluate different textures corresponding to different $D$-values by inserting the quark mass values at the $M_Z$-scale \cite{ak1jamin}, 
\[
m_u(M_Z)=1.7 MeV, m_c(M_Z)=0.62 GeV, m_t(M_Z)=171 GeV,
\]
\[
m_d(M_Z)=3 MeV, m_s(M_Z)=54 MeV, m_b(M_Z)=2.9 GeV,
\]
which for $D = C - A - B$ and $D' = C' - A' - B'$, gives
\begin{equation}\label{ak1nambyu0}
{\bf{L}_{(D=C-A-B)}} = A 
 \left (\begin{array}{rcl}
                  1 {\hspace{6mm}}      &{\hspace{2mm}}  0.9999{\hspace{10mm}}1.978\nonumber\\
                  0.9999                & {\hspace{6mm}} 1{\hspace{15mm}}1.978\nonumber\\
                  1.978{\hspace{2mm}}   &{\hspace{3mm}}1.978{\hspace{10mm}}3.9782
                      \end{array}
                \right)
\end{equation}
and 
\begin{equation}\label{ak1nambydd}
{\bf{L'}_{(D'=C'-A'-B')}}= A'
               \left(\begin{array}{rcl}
                  1{\hspace{6mm}}       &{\hspace{2mm}}  0.994{\hspace{10mm}}1.886\nonumber\\
                  0.994{\hspace{2mm}}   & {\hspace{6mm}} 1{\hspace{15mm}}1.886\nonumber\\
                  1.886{\hspace{2mm}}   &{\hspace{3mm}}  1.886{\hspace{10mm}}3.8795
    \end{array}
                \right)
\end{equation}
with $A = 28.7 GeV$ and $A' = 0.497 GeV$, which corresponds to 
a texture  
\begin{equation}\label{ak1nambydi}
{\bf{M}}\approx {\bf{C}} 
               \left(\begin{array}{rcl}
                  1{\hspace{8mm}}          &{\hspace{2mm}} 1-\varepsilon^3{\hspace{10mm}}2-\varepsilon\nonumber\\
        1-\varepsilon^3{\hspace{2mm}}       & {\hspace{6mm}} 1{\hspace{15mm}}2-\varepsilon\nonumber\\
2-\varepsilon{\hspace{2mm}}{\hspace{2mm}}   &{\hspace{3mm}}  2-\varepsilon{\hspace{10mm}}4-\varepsilon
    \end{array}
                \right)
\end{equation}

\noindent Instead of chosing a specific $D$-value, we introduce the notation $D=\pm k(C-A-B)$, and $t=\sqrt{1+8k^2}$, which gives the mass spectrum
$(A-B,[(A+B)(1+t)+C(1-t)]/2,[(A+B)(1-t)+C(1+t)]/2)$.
\\
The remaining flavour
symmetry is broken by
introducing phases. 
This gives the mass matrices 
${\bf{M}_u}$ and ${\bf{M}_d}$, for the up- and down-sectors, respectively:
\\
${\bf{M}_u}={\bf{\Omega}}^{\dagger}{\bf{L}}{\bf{\Omega}}$ and
${\bf{M}_d}={\bf{\Omega'}}^{\dagger}{\bf{L'}}{\bf{\Omega'}}$,
\\
\begin{equation}
{\bf{\Omega}}=\left(\begin{array}{rcl}
         1&{\hspace{2mm}}0{\hspace{6mm}}0\nonumber\\
         0&{\hspace{2mm}}e^{i\beta}{\hspace{4mm}}0\nonumber\\
         0&{\hspace{4mm}}0{\hspace{4mm}}e^{i\alpha}
                      \end{array}
                \right)\nonumber
{\hspace{4mm}}{\rm{and}}{\hspace{4mm}}{\bf{\Omega'}}=\left(\begin{array}{rcl}
         1&{\hspace{2mm}}0{\hspace{6mm}}0\nonumber\\
         0&{\hspace{2mm}}e^{i\mu}{\hspace{4mm}}0\nonumber\\
         0&{\hspace{4mm}}0{\hspace{4mm}}e^{i\rho}         
                      \end{array}
                \right),\nonumber
\end{equation}

\begin{equation}\label{ak1aa11}
{\bf{M}_u} = \left(\begin{array}{rcl}
         A{\hspace{7mm}} &  Be^{i\beta}{\hspace{11mm}}De^{i\alpha}\nonumber\\
             Be^{-i\beta} &{\hspace{6mm}}  A{\hspace{15mm}}
De^{i(\alpha-\beta)}\nonumber\\
         De^{-i\alpha}   &  De^{-i(\alpha-\beta)}{\hspace{9mm}}C
                      \end{array}
                \right)
\end{equation}

and
\begin{equation}\label{ak1aa22}
{\bf{M}_d} = \left(\begin{array}{rcl}
    A'{\hspace{7mm}}              & B'e^{i\mu}{\hspace{11mm}}D'e^{i\rho}\nonumber\\
        B'e^{-i\mu}                &{\hspace{6mm}}A'{\hspace{15mm}}D'e^{i(\rho-\mu)}\nonumber\\
    D'e^{-i\rho}                   &  D'e^{i(\mu-\rho)}{\hspace{9 mm}}C'
                      \end{array}
                \right)
\end{equation}
\\
with the mass spectra
$(A-B,[(A+B)(1+t)+C(1-t)]/2,[(A+B)(1-t)+C(1+t)]/2)$ with $D$ = $k(C-A-B)$, $t$ = $sqrt{1+8k^2}$, and
$(A'-B',[(A'+B')(1+t')+C'(1-t')]/2,[(A'+B')(1-t')+C'(1+t')]/2)$, 
$D'=k'(C'-A'-B')$ $t'=\sqrt{1+8k'^2}$.

The mass matrices (\ref{ak1aa11}) and (\ref{ak1aa22}) are diagonalized by the 
unitary matrices 

\begin{equation}\label{ak1di11}
U_u = \left(\begin{array}{rcl}
         c_{\eta}{\hspace{11mm}}   &                -s_{\eta}e^{i\beta}
{\hspace{16mm}}0\nonumber\\
 s_{\eta}c_{\gamma}{\hspace{7mm}}  &{\hspace{6mm}}c_{\eta}c_{\gamma}e^{i\beta}
{\hspace{15mm}}s_{\gamma}e^{i\alpha}\nonumber\\
s_{\eta}s_{\gamma}{\hspace{7mm}}   &{\hspace{6mm}}c_{\eta} s_{\gamma}e^{i\beta}
{\hspace{11mm}}-c_{\gamma}e^{i\alpha}
                      \end{array}
                \right)
\end{equation}

and

\begin{equation}\label{ak1di22}
U_d = \left(\begin{array}{rcl}
          c_{\phi}{\hspace{11mm}}   & -s_{\phi}e^{i\mu}
{\hspace{16mm}}0\nonumber\\
    s_{\phi}c_{\nu}{\hspace{7mm}}  &{\hspace{6mm}}c_{\phi}c_{\nu}e^{i\mu}
{\hspace{15mm}}s_{\nu}e^{i\rho}\nonumber\\
   s_{\phi}s_{\nu}{\hspace{7mm}}   &{\hspace{6mm}}c_{\phi} s_{\nu}e^{i\mu}
{\hspace{11mm}}-c_{\nu}e^{i\rho}
                      \end{array}
                \right)
\end{equation}
\\

\noindent where $\sin \eta \cos \eta=1/2$, $\cos {\phi} \sin{\phi}=1/2$.
The angles $\gamma$ and $\nu$ are given by $\tan 2 \gamma = - 2\sqrt{2}k = -\sqrt{t^2 - 1}$ and $\tan 2 \nu = - 2\sqrt{2}k' = -\sqrt{t'^2 - 1}$, which gives
\begin{eqnarray}\label{ak1gamma}
\sin ^2\gamma  =  \frac{t-1}{2t}, \cos ^2\gamma  =  \frac{t+1}{2t},{\hspace{2mm}}{\rm{or}} \nonumber\\
\sin ^2\gamma  =  \frac{t+1}{2t}, \cos ^2\gamma  =  \frac{t-1}{2t},{\hspace{2mm}}
\end{eqnarray}
And analogously for $\cos^2 \nu$, $\sin^2 \nu$.

\section{The mixing matrix}

With the diagonalizing matrices $U_u$ and $U_d$,
the mixing matrix $V = U_u U_d^{\dagger}$ takes the form
\begin{eqnarray}\label{ak1kmix}
V&=& \frac{1}{2}
\left(\begin{array}{rcl}
        (1+e^{i\theta}){}    & c_{\nu}(1-e^{i\theta})
{\hspace{28mm}}  s_{\nu}(1-e^{i\theta})\nonumber\\
 c_{\gamma}(1-e^{i\theta}){} &{\hspace{4mm}} c_{\nu}
c_{\gamma}(1+e^{i\theta})+ 2 s_{\gamma}s_{\nu}e^{i\psi}
{\hspace{7mm}}  c_{\gamma}s_{\nu}(1+e^{i\theta})-2s_{\gamma}c_{\nu}e^{i\psi}\nonumber\\
 s_{\gamma}(1-e^{i\theta}){} &{\hspace{4mm}}  c_{\nu} 
s_{\gamma}(1+e^{i\theta})-2c_{\gamma}s_{\nu}e^{i\psi}
{\hspace{7mm}} s_{\gamma}s_{\nu}(1+e^{i\theta})+2c_{\gamma}c_{\nu}e^{i\psi}
      \end{array}                
\right)\\
\\
&=& e^{i\frac{\theta}{2}} 
\left(\begin{array}{rcl}
                \cos \frac{\theta}{2}{\hspace{10mm}}& -i c_{\nu}\sin \frac{\theta}{2}{\hspace{21mm}}  
-i s_{\nu}\sin \frac{\theta}{2}\nonumber\\
-i c_{\gamma}\sin \frac{\theta}{2}{\hspace{7mm}}    &{\hspace{4mm}} c_{\nu}c_{\gamma}\cos \frac{\theta}{2}
+  s_{\gamma}s_{\nu}e^{i\Delta}
{\hspace{7mm}}  c_{\gamma}s_{\nu}\cos \frac{\theta}{2}-s_{\gamma}c_{\nu}e^{i\Delta}\nonumber\\
-i s_{\gamma}\sin \frac{\theta}{2}{\hspace{7mm}}    &{\hspace{4mm}}c_{\nu}s_{\gamma} \cos \frac{\theta}{2}-
c_{\gamma}s_{\nu}e^{i\Delta}{\hspace{7mm}} s_{\gamma}s_{\nu}\cos \frac{\theta}{2}+
c_{\gamma}c_{\nu}e^{i\Delta}\nonumber
      \end{array}
\right)
\end{eqnarray}
\\

\noindent where $\theta = \beta - \mu$, $\psi = \alpha - \rho$ and
$\Delta = \alpha - \rho - \theta/2$.
Introducing the notation 

\begin{eqnarray}
\cos \frac{\theta}{2}&&={\hspace{3mm}}c_1,{\hspace{16mm}} 
\sin \frac{\theta}{2}{\hspace{3mm}}={\hspace{3mm}}s_1\nonumber\\
 c_{\gamma}          &&={\hspace{3mm}}ic_2, {\hspace{20mm}} 
s_{\gamma}{\hspace{3mm}}={\hspace{3mm}}is_2 \nonumber\\
 c_{\nu}             &&={\hspace{3mm}}-ic_3,{\hspace{14mm}}  
s_{\nu}{\hspace{3mm}}={\hspace{3mm}}-is_3 \\
e^{i\Delta}         &&={\hspace{3mm}}-e^{i \delta}\nonumber,
\end{eqnarray}
and discarding the overall phase factor, the mixing matrix takes the form
\begin{equation}
V=\left(\begin{array}{rcl}
        c_1   &{\hspace{1mm}}-s_1c_3{\hspace{16mm}}-s_1s_3\nonumber\\
        s_1c_2&{\hspace{3mm}}c_1c_2c_3-s_2s_3e^{i\delta}{\hspace{6mm}}
        c_1c_2s_3+s_2c_3e^{i\delta}\\
        s_1s_2&{\hspace{3mm}}c_1s_2c_3+c_2s_3e^{i\delta}{\hspace{6mm}}
        c_1s_2s_3-c_2c_3e^{i\delta}\nonumber
               \end{array}
         \right)
\end{equation}

\subsection{The mass matrix elements}

From (\ref{ak1gamma}) we see that $t=1/(1-2\sin^2\gamma)$ or $t=1/(1-2\cos^2\gamma)$.
Introducing explicit numerical values\cite{ak1data} for the mixing matrix elements, 
$|V_{ud}|=0.9742$, $|V_{us}|=0.2253$, $|V_{ub}|=0.003437$, and $|V_{cd}|=0.2252$,  and $|V_{td}|=0.00862$,
we get $\cos \theta/2 =0.97428$, $\sin \nu = 0.0154$ and 
$\sin \gamma = 0.0382$. 

Inserting this into the expressions for $t$ gives
$t=\pm 1.0029$, and analogously
$t'=\pm 1.00047$, corresponding to $k=0.0269$ and $k'=0.01089$.
The measured mass values are not the same as the tree level mass spectra that one gets from the mass matrices, but using data on quark mass values still give a hint of the relative values of the matrix parameters.   

Inserting the quark mass values\cite{ak1jamin} at the $M_Z$-scale,
\[
m_u(M_Z)=1.7 MeV, m_c(M_Z)=0.62 GeV, m_t(M_Z)=171 GeV,
\]
\[
m_d(M_Z)=3 MeV, m_s(M_Z)=54 MeV, m_b(M_Z)=2.9 GeV,
\]
into the expressions

$A = {\hspace{2mm}}[2tm_u+m_c(t+1)+m_t(t-1)]/4$, $B = {\hspace{2mm}}[-2tm_u+m_c(t+1)+m_t(t-1)]/4$,...
and similarly for $A'$, $B'$,..and positive $t$ and $t'$,
gives
\begin{equation}\label{ak1aa33}
{\bf{M}_u} \approx B \left(\begin{array}{rcl}
1.004{\hspace{6mm}}     &{\hspace{9mm}}e^{i\beta}{\hspace{21mm}}-10.6e^{i\alpha}\nonumber\\
e^{-i\beta}{\hspace{6mm}} &{\hspace{14mm}}  1.004{\hspace{18mm}}
-10.6e^{i(\alpha-\beta)}\nonumber\\
  -10.6e^{-i\alpha}       &-10.6e^{-i(\alpha-\beta)}{\hspace{14mm}}395
                      \end{array}
                \right)
\end{equation}

\begin{equation}\label{ak1aa44}
{\bf{M}_d} \approx B' \left(\begin{array}{rcl}
1.12{\hspace{6mm}}     &{\hspace{12mm}}e^{i\mu}{\hspace{19mm}}-1.2e^{i\rho}\nonumber\\
e^{-i\mu}{\hspace{6mm}} &{\hspace{18mm}}  1.12{\hspace{18mm}}
-1.2e^{i(\rho-\mu)}\nonumber\\
  -1.2e^{-i\rho}       &{\hspace{2mm}}-1.2e^{-i(\rho-\mu)}{\hspace{12mm}}111
                      \end{array}
                \right) 
\end{equation}
\\
with $B=0.43 GeV$, $B'=25.8 MeV$, corresponding to the texture

\begin{equation}\label{ak1aaz}
|{\bf{M}}| \approx {\bf{C}} \left(\begin{array}{rcl}
 1+\delta{\hspace{12mm}} & {\hspace{13mm}} 1{\hspace{23mm}}1+x\nonumber\\
       1{\hspace{15mm}}  &{\hspace{10mm}}  1+\delta{\hspace{18mm}}1+x\nonumber\\
   1+x{\hspace{11mm}}    &{\hspace{10mm}}  1+x{\hspace{18mm}}1+y
                      \end{array}
                \right)
\end{equation}
where $x$ is negative and $y \gg 1$. 

\section{Four families}

An obvious way of extending the Standard Model is to introduce a fourth family of quarks and leptons,
the question of the number of families is one of the recurrent themes of discussion at the Bled workshops\cite{ak1bled}.

Fourth family\cite{ak1Kribs} leptons and quarks have been searched for at LEP, the Tevatron and LHC, giving the mass limits 
\[
m_L > 100.8{\hspace{2mm}} GeV\\
\]
\[
m_B > 372{\hspace{2mm}} GeV\\
\]
\[
m_T > 335,{\hspace{2mm}} GeV
\]
where $m_L$ is a charged lepton mass.
Moreover, an important characteristic of the fourth generation quark doublet is that the mass splitting between the heavy up- and down-quarks is constrained by electroweek precision tests to be small, likely less than $M_W$\cite{ak14eth}.

Playing the same game as in ({\ref{ak1namby}}) for four families, we get the matrix  
\begin{equation}\label{ak1nambww}
{\bf{M}} = \left(\begin{array}{rcl}
                  A   &  B{\hspace{4mm}}D{\hspace{4mm}}E\nonumber\\
                  B   &  A{\hspace{4mm}}D{\hspace{4mm}}E\\
                  D   &  D{\hspace{4mm}}X{\hspace{4mm}}E\nonumber\\
                  E   &  E{\hspace{4mm}}E{\hspace{4mm}}C 
                      \end{array}
                \right)
\end{equation}
This matrix has two redundant parameters, and

$\Tr({\bf{M}}$) = 2A+D+C,

$det({\bf{M}})=\frac{1}{24} [(\Tr{\bf{M}})^4+3(\Tr({\bf{M}}^2))^2+8(\Tr({\bf{M}}^3))\Tr({\bf{M}})-6\Tr({\bf{M}}^4)-6\Tr({\bf{M}}^2)(\Tr{\bf{M}})^2]$
= $(A-B)[(A+B)(CX-E^2)+4DE^2-2XE^2-2CD^2]$

For $X$ = $D$, we get
\begin{equation}\label{ak1nambyy}
{\bf{M}} = \left(\begin{array}{rcl}
                  A   &  B{\hspace{4mm}}D{\hspace{4mm}}E\nonumber\\
                  B   &  A{\hspace{4mm}}D{\hspace{4mm}}E\\
                  D   &  D{\hspace{4mm}}D{\hspace{4mm}}E\nonumber\\
                  E   &  E{\hspace{4mm}}E{\hspace{4mm}}C 
                      \end{array}
                \right)
\end{equation}
with $\Tr({\bf{M}}) = 2A+D+C$ and $det({\bf{M}}) = (A-B)(A+B-2D)(CD-E^2)$.
From the matrix invariants 
\begin{align*}
m_1m_2+&m_1m_3+m_1m_4+m_2m_3+m_2m_4+m_3m_4 \\
 &= [(\Tr({\bf{M}}))^2-\Tr({\bf{M}}^2)]/2
\\
m_1m_2m_3+&m_1m_2m_4+m_1m_3m_4+m_2m_3m_4 \\
&= [(\Tr({\bf{M}}))^3+2{\hspace{1mm}}\Tr({\bf{M}}^3)-3{\hspace{1mm}}\Tr{\bf{M}}{\hspace{1mm}}\Tr({\bf{M}}^2)]/6 
\end{align*}
we get the eigenvalues: 
\[
(m_1,m_2,m_3,m_4) =  (A-B, A+B-2D, (3D+C \pm \sqrt{(3D+C)^2+4E^2-4CD})/2),
\]
where $E^2 =2D^2+CD-(A+B)D$, and thus $det({\bf{M}}) = D(A-B)(A+B-2D)^2$. 

From these eigenvalues we infer $A = [m_2^2+2m_3m_4+m_1m_2]2m_2$, $B = [m_2^2+2m_3m_4-m_1m_2]/2m_2$, $C = [m_2m_3+m_2m_4-3m_3m_4]m_2$, $D = m_3m_4/m_2$
and like in the 3x3 case, we can insert mass values in the mass matrix elements, using 
\[
m_T = 311 GeV, m_B= 338 GeV,
\]
heavy quark masses based on new experimental limits on heavier quarks
are a little higher than the ones from The Particle Data Group\cite{ak14eth} of 2010,   
\[
m_T \lesssim 256 GeV, m_B \lesssim 128 GeV
\]
The resulting numerical matrix for the up-sector is
\begin{equation}\label{ak1nambup}
{\bf{M}} = D \left (\begin{array}{rcl}
                  1.00000362 &  1.00000360{\hspace{8mm}}1{\hspace{14mm}}1.73i\nonumber\\
                  1.00000360 &  1.00000362{\hspace{8mm}}1{\hspace{14mm}}1.73i\\
             1{\hspace{10mm}}& {\hspace{11mm}} 1{\hspace{14mm}}1{\hspace{14mm}}1.73i\nonumber\\
          1.73i{\hspace{8mm}}&  {\hspace{9mm}}1.73i{\hspace{8mm}}1.73i{\hspace{4mm}}-2.99 
                      \end{array}
                \right) \approx
\end{equation}

\begin{equation}\label{ak1nambup2}
{\bf{M}} = D \left (\begin{array}{rcl}
                  1.00000362 &  1.00000360{\hspace{8mm}}1{\hspace{14mm}}1.73i\nonumber\\
                  1.00000360 &  1.00000362{\hspace{8mm}}1{\hspace{14mm}}1.73i\\
             1{\hspace{10mm}}& {\hspace{11mm}} 1{\hspace{14mm}}1{\hspace{14mm}}1.73i\nonumber\\
          1.73i{\hspace{8mm}}&  {\hspace{9mm}}1.73i{\hspace{8mm}}1.73i{\hspace{4mm}}-2.99 
                      \end{array}
                \right) \approx
\end{equation}
with $D$ = $85775.80 GeV.$

Similarly for the down-sector
\begin{equation}\label{ak1nambdown}
{\bf{M'}} = D \left (\begin{array}{rcl}
                  1.0000016  &  1.0000014{\hspace{10mm}}1{\hspace{14mm}}1.726i\nonumber\\
                  1.0000014  &  1.0000016{\hspace{10mm}}1{\hspace{14mm}}1.726i\\
             1{\hspace{10mm}}& {\hspace{11mm}} 1{\hspace{14mm}}1{\hspace{14mm}}1.726i\nonumber\\
         1.726i{\hspace{6mm}}&  {\hspace{7mm}}1.726i{\hspace{8mm}}1.726i{\hspace{4mm}}-2.98 
                      \end{array}
                \right) 
\end{equation}
with $D'$ = 17964.074 GeV.

Both these matrices display this texture  
\begin{equation}\label{ak1nambq}
{\bf{\cal{M}}} = {\bf{C}} \left (\begin{array}{rcl}
                    1+\delta &  1+\gamma{\hspace{10mm}}1{\hspace{14mm}}i\varepsilon \nonumber\\
                    1+\gamma &  1+\delta{\hspace{10mm}}1{\hspace{14mm}}i\varepsilon\\
              1{\hspace{4mm}}& {\hspace{5mm}} 1{\hspace{12mm}}1{\hspace{14mm}}i\varepsilon \nonumber\\
  i \varepsilon{\hspace{3mm}}& {\hspace{6mm}} i \varepsilon{\hspace{10mm}}i \varepsilon{\hspace{9mm}}- \varepsilon^2
                      \end{array}
                \right) 
\end{equation}
where $\varepsilon \approx \sqrt{3}$ and both $\gamma$ and $\delta$ are very small, $\sim$ $10^{-6}$. 
The phases remains to be introduced, and the mixing matrix to be determined.


\begin{thebibliography}{99}

\bibitem{ak1demo} H. Harari, H. Haut, and J. Weyers,
  Phys. Lett. \textbf{78B}   (1978) 459.

\bibitem{ak1koide} T. Goldman and G. J. Stephenson Jr.,
  Phys. Rev. \textbf{D24}  (1981) 236,
and Y. Koide, Phys. Rev. Lett. \textbf{47}  (1981) 1241.

\bibitem{ak1Fritzsch}  H. Fritzsch, Phys.Lett. \textbf{73B} (1978) 317. 

\bibitem{ak1data} K. Nakamura et al. (2010). "Review of Particles Physics: The CKM Quark-Mixing Matrix". Journal of Physics G 37 (75021): 150. http://pdg.lbl.gov/2010/reviews/rpp2010-rev-ckm-matrix.pdf.

\bibitem{ak1jamin} M. Jamin. ICREA and IFAE. Universitat Aut`onoma de Barcelona. Quark Masses. Granada (2006).

\bibitem{ak1bled}
Proceedings of the Bled Workshops, hep-ph/9905357v1; hep-ph/0301029v1; hep-ph/0412208; hep-ph/0612250.

\bibitem{ak1Kribs} 
G. D. Kribs, T. Plehn, M. Spannowsky, T.M. Tain, Phys. Rev. \textbf{D 76}  (2007)  075016.

\bibitem{ak14th} D. Atwood, S. Kumar Gupta and A. Soni, hep-ph/1104.3874v3 (2011), X. Ruana,b , Z. Zhanga, hep-ph/1105.1634v2 (2011).

\bibitem{ak14eth} K. Nakamura et al. [Particle Data Group],
  J. Phys. \textbf{G  37}  (2010) 075021 and 2011 partial update for the 2012 edition. 

\end{thebibliography}
\end{document}